\documentclass[english]{scrartcl}
\usepackage{pc_math}
\usepackage[comma,authoryear]{natbib} 

\usepackage[x11names]{xcolor}
\definecolor{firebrick}{rgb}{.698,.133,.133}
\usepackage[colorlinks=true,allcolors=firebrick,bookmarks=false]{hyperref}

\usepackage{authblk}

\title{Divergence, Entropy, Information}
\subtitle{An Opinionated Introduction to Information Theory}

\author{
	Philip S. Chodrow\thanks{
		\texttt{pchodrow@mit.edu}
	}}
\affil{Operations Research Center and Laboratory for Information and Decision Systems \\

Massachusetts Institute of Technology
\protect\\ 77 Massachusetts Avenue, Cambridge, MA 02139 USA}

\date{\today}

\begin{document}
\setkomafont{disposition}{\mdseries\rmfamily}

\maketitle

\abstract{
	Information theory is a mathematical theory of learning with deep connections with topics as diverse as artificial intelligence, statistical physics, and biological evolution. 
	Many primers on information theory paint a broad picture with relatively little mathematical sophistication, while many others develop specific application areas in detail. 
	In contrast, these informal notes aim to outline some elements of the information-theoretic ``way of thinking,'' by cutting a rapid and interesting path through some of the theory's foundational concepts and results. 
	They are aimed at practicing systems scientists who are interested in exploring potential connections between information theory and their own fields. 
	The main mathematical prerequisite for the notes is comfort with elementary probability, including sample spaces, conditioning, and expectations.  

	We take the Kullback-Leibler divergence as our most basic concept, and then proceed to develop the entropy and mutual information. 
	We discuss some of the main results, including the Chernoff bounds as a characterization of the divergence; Gibbs' Theorem; and the Data Processing Inequality. 
	A recurring theme is that the definitions of information theory support natural theorems that sound ``obvious'' when translated into English.
	More pithily, ``information theory makes common sense precise.''
	Since the focus of the notes is not primarily on technical details, proofs are provided only where the relevant techniques are illustrative of broader themes. 
	Otherwise, proofs and intriguing tangents are referenced in liberally-sprinkled footnotes. 
	The notes close with a highly nonexhaustive list of references to resources and other perspectives on the field. 
}

\newpage
\tableofcontents
\newpage

\section{Why Information Theory?}
	
	Briefly, information theory is a mathematical theory of learning with rich connections to physics, statistics, and biology. 
	Information-theoretic methods quantify \emph{complexity} and \emph{predictability} in systems, and make precise how observing one feature of a system assists in predicting other features. 
	Information-theoretic thinking helps to structure algorithms; describe processes in natural and engineered systems; and draw surprising connections between seemingly disparate fields. 

	Formally, information theory is a subfield of probability, the mathematical study of uncertainty and randomness.
	Information theory is distinctive in its emphasis on properties of probability distributions that are independent of how those distributions are represented. 
	Because of this representation-independence, information-theoretic quantities often have claim to be the most ``fundamental'' properties of a system or problem, governing its complexity, learnability, and intrinsic randomness. 

	In the original formulation of \cite{Shannon1948}, information theory is a theory of \emph{communication}: specifically, the transmission of a signal of some given complexity over an unreliable channel, such as a telephone line corrupted by a certain amount of white noise. 
	Here we will emphasize a slightly different role for information theory, as a theory of \emph{learning} more generally. 
	This emphasis is consistent with the original formulation, since the communication problem may be viewed as the problem of the message receiver learning the intent of the sender based on the potentially corrupted transmission. 
	However, the emphasis on learning allows us to more easily glimpse some of the rich connections of information theory to other disciplines. 
	Special consideration in these notes will be given to statistical motivations for information-theoretic concepts. 
	Theoretical statistics is the mathematical design of methods for learning from data; information-theoretic considerations determine when such learning is possible and to what extent. 
	We will close with a connection to physics; some connections to biology are cited in the references. 

\section{Why Not Start with Entropy?}
	
	\emph{Entropy} is easily the information-theoretic concept with the widest popular currency, and many expositions take entropy as their starting point. 
	We will, however, choose a different point of departure and derive entropy along the way. 
	Our primary object is the \emph{Kullback-Leibler (KL) divergence} between two distributions, also called in some contexts the \emph{relative entropy}, \emph{relative information}, or \emph{free energy}.\footnote{For the remainder of these notes I'll stick with ``divergence'' -- though there are many other interesting objects called ``divergences'' in mathematics, we won't be discussing any of them here, so no confusion should arise.}
	Why start with the divergence? 
	Well, there's a simple reason -- while we'll focus on discrete random variables here, we'd like to develop a theory that, wherever possible, applies to continuous random variables as well. 
	The divergence is well-defined for both discrete random variables and continuous ones; that is, if $p$ and $q$ are two continuous distributions satisfying certain regularity properties, then $d(p,q)$ is a uniquely determined, nonnegative real number whether $p$ and $q$ are discrete or continuous. 
	In contrast, the natural definition of entropy (so called \emph{differential entropy}) for continuous random variables has two bad behaviors. 
	First, it can be negative, which is undesirable for a measure of uncertainty. 
	Second, and arguably worse, the differential entropy is not even uniquely defined.
	There are multiple ways to describe the same continuous distribution -- for example, the following three distributions are the same: 
	\begin{enumerate}
		\item ``The Gaussian distribution with mean 0 and variance 1.''
		\item ``The Gaussian distribution with mean 0 and standard deviation 1.''
		\item ``The Gaussian distribution with mean 0 and second moment equal to 1.''
	\end{enumerate}
	Technically, the act of switching from one of these descriptions to another can be viewed as a smooth change of coordinates\footnote{I.e. a \emph{diffeomorphism}: smooth, invertible functions on coordinate space whose inverses are also smooth.} in the space of distribution parameters. 
	For example, we move from the first description to the second by changing coordinates from $(\mu, \sigma^2)$ to $(\mu, \sigma)$, which we can do by applying the function $f(x,y) = (x, \sqrt{y})$. 
	Regrettably, the differential entropy is not invariant under such coordinate changes -- change the way you describe the distribution, and the differential entropy changes as well. 
	This is undesirable. 
	The foundations of our theory should be independent of the contingencies of how we describe the distributions under study. 
	The divergence passes this test in both discrete and continuous cases; the differential entropy does not.\footnote{It is possible to define alternative notions of entropy that attempt to skirt these issues; however, they have their own difficulties. \url{https://en.wikipedia.org/wiki/Limiting_density_of_discrete_points}} 
	Since we can define the entropy in terms of the divergence in the discrete case, we'll start with the divergence and derive the entropy along the way. 

\section{Introducing the Divergence}
	It is often said that the divergence $d(p,q)$ between distributions $p$ and $q$ measures how ``surprised'' you are if you think the state of the world is $q$ but then measure it to be $p$. 
	However, this idea of surprise isn't always explained or made precise. 
	To motivate the KL divergence, we'll start from a somewhat unusual beginning -- the Chernoff bounds -- that makes exact the role that the divergence plays in governing how surprised you ought to be. 

	Let's begin with a simple running example. 
	You are drawing from an (infinite) deck of standard playing cards, with four card suits $\{\spadesuit, \clubsuit, \color{firebrick}\diamondsuit, \color{firebrick}\heartsuit\color{black}\}$ and thirteen card values $\{1,\ldots,13\}$. 
	We'll view the sets of possible values as \emph{alphabets}: $\mathcal{X} = \{\spadesuit, \clubsuit, \color{firebrick}\diamondsuit, \color{firebrick}\heartsuit\color{black}\}$ is the alphabet of possible suits, and $\mathcal{Y} = \{1,\ldots,13\}$ the alphabet of possible values. 
	We'll let $X$ and $Y$ be the corresponding random variables, so for each realization, $X \in \mathcal{X}$ and $Y \in \mathcal{Y}$.  

	Suppose that I have a prior belief that the distribution of suits in the deck is uniform. 
	My belief about the suits can be summarized by a vector $q = (\frac{1}{4},\frac{1}{4}, \frac{1}{4},\frac{1}{4})$. 
	It's  convenient to view $q$ as a single point in the \emph{probability simplex} $\mathcal{P}^{\mathcal{X}}$ of all valid probability distributions over $\mathcal{X}$. 
	\begin{dfn}[Probability Simplex]
		For any finite alphabet $\mathcal{X}$ with $\abs{X} = m$, the \textbf{probability simplex} $\mathcal{P}^\mathcal{X}$ is the set 
		\begin{equation*}
			\mathcal{P}^\mathcal{X} \triangleq \left\{ q \in \R^m \;|\; \sum_{i}q_i = 1,\; q_i \geq 0 \quad \forall i  \right\}\;.
		\end{equation*}
	\end{dfn}
	\begin{remark}
		It's helpful to remember that $\mathcal{P}^\mathcal{X}$ is an $m-1$-dimensional space; the ``missing'' dimension is due to the constraint $\sum_i q_i = 1$. 
		When $m = 3$, $\mathcal{P}^{\mathcal{X}}$ is an equilateral triangle; when $m = 4$ a tetrahedron, and so on. 
	\end{remark}

	If $q$ is your belief, you would naturally expect that, if you drew enough cards, the observed distribution of suits would be ``close'' to $q$, and that if you could draw infinitely many cards, the distribution would indeed converge to $q$. 
	Let's make this precise: define $\hat{p}_n \in \mathcal{P}^{\mathcal{X}}$ to be the distribution of suits you observe after pulling $n$ cards. 
	It's important to remember that $\hat{p}$ is a random vector, which changes in each realization. 
	But it would be reasonable to expect that $\hat{p} \rightarrow q$ as $n\rightarrow \infty$, and indeed this is true almost surely (with probability 1) according to the Strong Law of Large Numbers, \emph{if $q$ is in fact the true distribution of cards in the deck}. 

	But what happens if you keep drawing cards and the observed distribution $\hat{p}_n$ is much different than your belief $q$? 
	Then you would justifiably be surprised, and your ``level of surprise'' can be quantified by the probability of observing $\hat{p}_n$ if the true distribution were $q$, which I'll denote $\prob(\hat{p}_n ; q)$. 
	We'd naturally expect $\prob(\hat{p}_n ; q)$  to become small when $n$ grows large. 
	Indeed, there is a quite strong result here -- $\prob(\hat{p}_n ; q)$ decays exponentially in $n$, with a very special exponent.

	\begin{dfn}[Kullback-Leibler Divergence]
		For $p,q \in \mathcal{P}^{X}$, the \textbf{Kullback-Leibler (KL) divergence} of $q$ from $p$ is 
		\begin{equation*}
			d(p,q) \triangleq \sum_{x \in \mathcal{X}} p(x) \log \frac{p(x)}{q(x)}\;,
		\end{equation*}
		where we are using the conventions that $\log \infty = \infty$, $\log 0 = -\infty$, $0 / 0 = 0$, and $0 \times \infty = 0$. 
	\end{dfn}
	\begin{thm}[Chernoff Bounds] \label{thm:chernoff}
		Suppose that the card suits are truly distributed according to $p \neq q$. Then, 
		\begin{equation*}
			\frac{e^{-nd(p,q)}}{(n+1)^m}\leq \prob(\hat{p}_n;q) \leq e^{-n d(p,q)}\;.
		\end{equation*}
	\end{thm}

	So, the probability of observing $\hat{p}_n$ when you thought the distribution was $q$ decays exponentially, with the exponent given by the divergence of your belief $q$ from the true distribution $p$. 
	Another way to say this: ignoring non-exponential factors, 
	\begin{equation*}
		- \frac{1}{n} \log \prob(\hat{p}_n;q)\cong d(p,q)\;,
	\end{equation*}
	that is, $d(p,q)$ is the minus the log of your average surprise per card drawn. 
	The Chernoff bounds thus provide firm mathematical content to the idea that the divergence measures surprise. 

	To make this result concrete, suppose  I start with the belief that the deck is uniform over suits. 
	So, my belief is $q = (\frac{1}{4},\frac{1}{4}, \frac{1}{4},\frac{1}{4})$ over the alphabet $\{\spadesuit, \clubsuit, \color{firebrick}\diamondsuit, \color{firebrick}\heartsuit\color{black}\}$. 
	Unbeknownst to me, you have removed all the black cards from the deck, which therefore has true distribution $p = \left(0,0, \frac{1}{2}, \frac{1}{2} \right)$. 
	I draw 100 cards and record the suits. 
	How surprised am I by the suit distribution I observe? 
	The divergence between my belief and the true deck distribution is $d(p,q) \approx 0.69$, so Theorem \ref{thm:chernoff} states that the dominating factor in the probability of my observing an empirical distribution close to $p$ in 100 draws is $e^{-0.69 \times 100 } \approx 10^{-30}$. I am quite surprised indeed! 

	Can you have ``negative surprise?'' 
	Gibbs' inequality states that the answer is no: 
	\begin{thm}[Gibbs' Inequality] \label{thm:gibbs}
		For all $p,q \in \mathcal{P}^\mathcal{X}$, it holds that $d(p,q) \geq 0$, with equality iff $p = q$. 
	\end{thm}
	In words, you can never have ``negative surprise,'' and you are only unsurprised if you what you observed is exactly what you expected. 
	\begin{proof}
		There are lots of ways to prove Gibbs' Inequality; here's one with Lagrange multipliers.
		Fix $p \in \mathcal{P}^\mathcal{X}$. 
		We'd like to show that the problem 
		\begin{equation*}
			\min_{q \in \mathcal{P}^\mathcal{X}} d(p,q) 
		\end{equation*}
		has value 0 and that this value is achieved at the unique point $q = p$. 
		We need two gradients: the gradient of $d(p,q)$ with respect to $q$ and the gradient of the implicit constraint $g(q) \triangleq \sum_{x \in \mathcal{X}} q(x) = 1$. 
		The former is 
		\begin{align*}
			\nabla_q d(p,q) &= \nabla_q \left[ \sum_{x \in \mathcal{X}} p(x) \log \frac{p(x)}{q(x)} \right] \\ 
							&= - \nabla_q \left[ \sum_{x \in \mathcal{X}} p(x) \log q(x) \right] \\
							&= - p \oslash q,
		\end{align*}
		where $p \oslash q$ is the elementwise quotient of vectors, and where we recall the convention $0 / 0 = 0$. 
		On the other hand, 
		\begin{equation*}
			\nabla_q g(q) = \mathbf{1}, 
 		\end{equation*}
 		the vector whose entries are all unity. 
 		The method of Lagrange multipliers states that we should seek $\lambda \in \R$ such that 
 		\begin{equation*}
 			\nabla_qd(p,q) = \lambda \nabla_q(g_q), 
 		\end{equation*}
 		or 
 		\begin{equation*}
 			- p \oslash q = \lambda \mathbf{1},	
 		\end{equation*}
 		from which it's easy to see that the only solution is $q = p$ and $\lambda = -1$. 
 		It's a quick check that the corresponding solution value is $d(p,p) = 0$, which completes the proof. 
	\end{proof}

	\begin{remark}
		Theorem \ref{thm:gibbs} is the primary sense in which $d$ behaves ``like a distance'' on the simplex $\mathcal{P}$. 
		On the other hand, $d$ is unlike a distance in that it is not symmetric and does not satisfy the triangle inequality.\footnote{In fact, $d$ is related to a ``proper'' distance metric on $\mathcal{P}^\mathcal{X}$, which is usually called the Fisher Information Metric and is the fundamental object of study in the field of information geometry \citep{Amari2000}.}
	\end{remark}

	Let's close out this section by noting one of the many connections between the divergence and classical statistics. 
	\emph{Maximum likelihood estimation} is a fundamental method of modern statistical practice; tools from linear regression to neural networks may be viewed as likelihood maximizers. 
	The divergence offers a particularly elegant formulation of maximum likelihood estimation: \textbf{likelihood maximization is the same as divergence minimization}. 
	Let $\theta$ be some statistical parameter, which may be multidimensional; for example, in the context of fitting normal distributions, we may have $\theta = (\mu, \sigma^2)$; in the context of regression, $\theta$ may be the regression coefficients $\beta$. 
	Let $p_X^{\theta}$ be the probability distribution over $X$ with parameters $\theta$. 
	Let $\{x_1, \ldots,x_n\}$ be a sequence of i.i.d. observations of $X$. 
	Maximum likelihood estimation encourages us to find the parameter $\theta$ such that 
	\begin{equation*}
		\theta^* = \argmax_\theta \prod_{i = 1}^n p_{X}^\theta(x_i)\;, 
	\end{equation*}
	i.e. the parameter value that makes the data most probable. To express this in terms of the divergence, we need just one more piece of notation: let $\hat{p}_X$ be the empirical distribution of observations of $X$. 
	Then, it's a slightly involved algebraic exercise to show that the maximum likelihood estimation problem can also be written 
	\begin{equation*}
		\theta^* = \argmin_\theta d(\hat{p}_X, p_X^\theta). 
	\end{equation*}
	This is rather nice  -- maximum likelihood estimation consists in making the parameterized distribution $p_X^\theta$ as close as possible to the observed data distribution $\hat{p}_X$ in the sense of the divergence.\footnote{This result is another hint at the beautiful geometry of the divergence: the operation of minimizing a distance-like measure is often called ``projection.'' 
	Maximum likelihood estimation thus consists in a kind of statistical projection.}
	
\section{Entropy}
	After having put it off for a while, let's define the Shannon entropy. 
	If we think about the divergence as a (metaphorical) distance, the entropy measures how close a distribution is to the uniform. 

	\begin{dfn}[Shannon Entropy] \label{def:entropy}
		The Shannon entropy of $p \in \mathcal{P}^{X}$ is 
		\begin{equation*}
			H(p) \triangleq -\sum_{x \in \mathcal{X}} p(x) \log p(x) \;.
		\end{equation*}
		When convenient, we will also use the notation $H(X)$ to refer to the entropy of a random variable $X$ distributed according to $p$. 
	\end{dfn}
	\begin{thm}\label{thm:entropy}
		The Shannon entropy is related to the divergence according to the formula
		\begin{equation} \label{eq:entropy}
			H(p) = \log m - d(p, u)\;,
		\end{equation}
		where $m = \abs{\mathcal{X}}$ is the size of the alphabet $\mathcal{X}$ and where $u$ is the uniform distribution on $\mathcal{X}$.\footnote{Here is as good a place as any to note that for discrete random variables, the divergence can also be defined in terms of the entropy. Technically, $d$ is the \emph{Bregman divergence} induced by the Shannon entropy, and can be characterized by the equation 
		\begin{equation*}
			d(p,q) =  - \left[H(p) - H(q) - \innerprod{\nabla_q H, p - q}\right]\;.
		\end{equation*}
		Intuitively, $d(p,q)$ is minus the approximation loss associated with estimating the difference in entropy between $p$ and $q$ using a first-order Taylor expansion centered at $q$. This somewhat artificial-seeming construction turns out to lead in some very interesting directions in statistics and machine learning.}  
	\end{thm}
	\begin{remark}
		This formula makes it easy to remember the entropy of the uniform distribution -- it's just $\log m$, where $m$ is the number of possible choices. 
		If we are playing a game in which I draw a card from the infinite deck, the suit of the card is uniform, and the entropy of the suit distribution is therefore $H(X) = \log 4 = 2\log 2$. 
	\end{remark}
	\begin{remark}
		In words, $d(p, u)$ is your surprise if you thought the suit distribution was uniform and then found it was in fact $p$. 
		If you are relatively unsurprised, then $p$ is very close to uniform. 
		Indeed, Gibbs' inequality (Theorem \ref{thm:gibbs}) immediately implies that $H(p)$ assumes its largest value of $\log m$ exactly when $p = u$. 
	\end{remark}	
	\begin{remark}
		Theorem \ref{thm:entropy} provides one useful insight into why the Shannon entropy does not generalize naturally to continuous distributions.
		Whereas equation \eqref{eq:entropy} expresses the entropy in terms of  the uniform distribution on $\mathcal{X}$, there can be no analogous formula for continuous random variables on $\R$ because there is no uniform distribution on $\R$. 
	\end{remark}

	\subsection*{A Bayesian Interpretation of Entropy}
		The construction of the entropy in terms of the divergence is fairly natural.
		We use the divergence to measure how close $p$ is to the uniform distribution, flip the sign so that high entropy distributions are more uniform, and add a constant term to make the entropy nonnegative. 
		This formulation of the entropy turns out to have another interesting characterization in the context of Bayesian prediction. 
		In Bayesian prediction, I will pull a single card from the deck. 
		Before I do, I ask you to provide a distribution $p$ over the alphabet $\{\spadesuit, \clubsuit, \color{firebrick}\diamondsuit, \color{firebrick}\heartsuit\color{black}\}$ representing your prediction about the suit of the card I pulled. 
		As examples, you can choose $p = (1,0,0,0)$ if you are certain that the suite will be $\spadesuit$, or $p = (\frac{1}{4}, \frac{1}{4}, \frac{1}{4}, \frac{1}{4})$ to express maximal ignorance. 
		After you guess, I pull the card, obtaining a sample $x \in \mathcal{X}$, and reward you based on the quality of your prediction relative to the outcome $x$. 
		I do this based on a loss function $f(p,x)$; after your guess I give you $f(p,x)$ dollars, say. 
		If we assume that my aim is to encourage you to (a) report your true beliefs about the deck and (b) reward you based only on what happened (i.e. not on what could have happened), then there is essentially only one appropriate loss function $f$, which turns out to be closely related to the entropy.
		More formally, 

		\begin{dfn}
			A loss function $f$ is \emph{proper} if, for any alphabet $\mathcal{Y}$ and random variable $Y$ on $\mathcal{Y}$, 
			\begin{equation*}
				p_{X|Y = y} = \argmin_{p \in \mathcal{P}^\mathcal{X}} \E[f(p, x)|Y = y].
			\end{equation*}
		\end{dfn}
		\begin{remark}
			In this definition, it's useful to think of $Y$ as some kind of ``side information'' or ``additional data.'' 
			For example, $Y$ could be my telling you that the card I pulled is a red card, which could influence your predictive distribution. 
			When $f$ is proper, you have an incentive to factor that into your predictive distribution. 
			While it may feel that ``of course'' you should factor this in, not all loss functions encourage you to do so. 
			For example, if $f$ is constant, then you have no incentive to use $Y$ at all, since each guess is just as good as any other. 
			A proper loss function guarantees that you can maximize your payout (minimize your loss) by completely accounting for all available data when forming your prediction, which should therefore be $p_{X|Y = y}$. Thus, a proper loss function ensures that the Bayesian prediction game is ``honest''. 	
		\end{remark}	
		\begin{dfn}
			A loss function $f$ is \emph{local} if $f(p,x) = \psi(p, p(x))$ for some function $\psi$. 
		\end{dfn}
		\begin{remark}
			The function $f$ is local iff $f$ can be written as a function only of my prediction $p$ and how much probabilistic weight I put on the event that actually occurred -- not events that ``could have happened'' but didn't. Thus, a proper loss function ensures that the Bayesian prediction game is ``fair.''
		\end{remark}

		Somewhat amazingly, the log loss function given by $f(p,x) = -\log p(x)$ is the \emph{only} loss function that is both proper and local (both honest and fair), up to an affine transformation.  
		\begin{thm}[Uniqueness of the Log-Loss]\label{thm:log_loss_uniqueness}
			Let $f$ be a local and proper reward function. Then, $f(p, x) = A \log p(x) + B$ for some constants $A < 0$ and $B \in \R$.  
		\end{thm}
		Without loss of generality, we'll take $A = -1$ and $B = 0$. 
		The entropy in this context occurs as the expected log-loss when you know the distribution of suits in the deck. 
		If you know, say, that the proportions in the deck are $p = (\frac{1}{4}, \frac{1}{4}, \frac{1}{4}, \frac{1}{4})$ and need to formulate your predictive distribution, Theorem \ref{thm:log_loss_uniqueness} implies that your best guess is just $p$, since you have no additional side information. 
		Then....
		\begin{dfn}[Entropy, Bayesian Characterization]
			The \textbf{(Shannon) entropy} of $p$ is your minimal expected loss when playing the Bayesian prediction game in which the true distribution of suits is $p$. 
		\end{dfn}
		\begin{remark}
			To see that this definition is consistent with the one we saw before, we can simply compute the expectation: 
			\begin{align*}
				\E[f(p,X)] &= \E[-\log p(X)] \\ 
				           &= -\sum_{x \in \mathcal{X}} p(x) \log p(x),
			\end{align*}
			which matches Definition \ref{def:entropy}. 
			The second inequality follows from the fact that, if you are playing optimally, $p$ is both the true distribution of $X$ and your best predictive distribution. 
		\end{remark}

\section{Conditional Entropy}
	The true magic of probability theory is conditional probabilities, which formalize the idea of learning: $\prob(A|B)$ represents my best belief about $A$ given what I know about $B$. 
	While the Shannon entropy itself is quite interesting, information theory really starts becoming a useful framework for thinking probabilistically when we formulate the conditional entropy, which encodes the idea of learning as a process of uncertainty reduction. 
	
	In this section and the next, we'll need to keep track of multiple random variables and distributions. 
	To fix notation, we'll let $p_X \in \mathcal{P}^\mathcal{X}$ be the distribution of a discrete random variable $X$ on alphabet $\mathcal{X}$, $p_Y \in \mathcal{P}^{\mathcal{Y}}$ the distribution of a discrete random variable $Y$ on alphabet $\mathcal{Y}$, and $p_{X,Y} \in \mathcal{P}^{\mathcal{X} \times \mathcal{Y}}$ their joint distribution on alphabet $\mathcal{X} \times \mathcal{Y}$. 
	Additionally, we'll denote the product distribution of marginals as $p_{X} \otimes p_{Y} \in \mathcal{P}^{\mathcal{X} \times \mathcal{Y}}$; that is, $(p_{X} \otimes p_Y)(x,y) = p_X(x) p_Y(y)$. 
	
	\begin{dfn}[Conditional Entropy]
		The \textbf{conditional entropy} of $X$ given $Y$ is 
		\begin{equation*}
			H(X|Y) \triangleq \sum_{x, y \in \mathcal{X} \times \mathcal{Y}} p_{X,Y}(x,y) \log p_{X|Y}(x|y)\;.
		\end{equation*}
	\end{dfn}
	\begin{remark}
		It might seem as though $H(X|Y)$ ought to be defined as 
		\begin{equation*}
			\tilde{H}(X|Y) = \sum_{x, y \in \mathcal{X} \times \mathcal{Y}} p_{X|Y}(x|y) \log p_{X|Y}(x|y)\;,
		\end{equation*}
		which looks more symmetrical. 
		However, a quick think makes clear that this definition isn't appropriate, because it doesn't include any information about the distribution of $Y$. 
		If $Y$ is concentrated around some very informative (or uninformative) values, then $\tilde{H}$ won't notice that some values of $Y$ are more valuable than others. 
	\end{remark}
	In the framework of our Bayesian interpretation of the entropy above, the conditional entropy is your expected reward in the guessing game assuming you receive some additional side information. 
	For example, consider playing the suit-guessing game in the infinite deck of cards. 
	Recall that the suit distribution is uniform, with entropy $H(X) = H(u) = 2 \log 2$. 
	Suppose now that you get side information -- when I draw the card from the deck, before I ask you to guess the suit, I tell you the color (black or red).
	Since for each color there are just two possible suits, the entropy decreases. 
	Formally, if $X$ is the suit and $Y$ the color, it's easy to compute that $H(X|Y) = \log 2$.
	Comparing to our previous calculation that $H(X) = 2\log2$, we see that knowing the color reduces your uncertainty by half. 

	The conditional entropy is somewhat more difficult to express in terms of the divergence, but it does have a useful relationship with the (unconditional) entropy. 
	\begin{thm}
		The conditional entropy is related to the unconditional entropy as 
		\begin{equation*}
			H(X|Y) = H(X,Y) - H(Y),
		\end{equation*}
		where $H(X,Y)$ is the entropy of the distribution $p_{X,Y}$.
	\end{thm}
	\begin{remark}
		This theorem is easy to remember, because it looks like what you get by recalling the definition of the conditional probability and taking logs: 
		\begin{equation*}
			p_{X|Y}(x|y) = \frac{p_{X,Y}(x,y)}{p_{Y}(y)}. 
		\end{equation*}
		Indeed, take logs and compute the expectations over $X$ and $Y$ to prove the theorem directly.  
		Another way to remember this theorem is to just say it out: the uncertainty you have about $X$ given that you've been told $Y$ is equal to the uncertainty you had about both $X$ and $Y$, less the uncertainty that was resolved when you learned $Y$. 
	\end{remark}
	From this theorem, it's a quick use of Gibbs' Inequality to show:
	\begin{thm}[Side Information Reduces Uncertainty] \label{thm:entropy_reduction}
		\begin{equation*}
			H(X|Y) \leq H(X). 
		\end{equation*}
	\end{thm}
	That is, knowing $Y$ can never make you more uncertain about $X$, only less. 
	This makes sense -- after all, if $Y$ is not actually informative about $X$, you can just ignore it. 

	Theorem \ref{thm:entropy_reduction} implies that $H(X) - H(X|Y) \geq 0$. 
	This difference quantifies how much $Y$ reduces uncertainty about $X$; if $H(X) - H(X|Y) = 0$, for example, then $H(X|Y) = H(X)$ and it is natural to say that $Y$ ``carries no information'' about $X$. 
	We encode the idea of information as uncertainty reduction in the next section. 
	
\section{Information Three Ways}
	
	Thus far, we've seen two concepts -- divergence and entropy -- that play fundamentals role in information theory. 
	But neither of them exactly resemble an idea of ``information,'' so how does the theory earn its name? 
	Our brief note at the end of the last section suggests that we think about information as a relationship between two variables $X$ and $Y$, in which knowing $Y$ decreases our uncertainty (entropy) about $X$. 
	As it turns out, the idea of information that falls out of this motivation is a remarkably useful one, and can be formulated in many interesting and different ways. 
	Let's start by formalizing this notion:
	\begin{dfn}[Mutual Information]
		The \textbf{mutual information} $I(X,Y)$ in $Y$ about $X$ is 
		\begin{equation*}
			I(X,Y) \triangleq H(X) - H(X|Y). 
		\end{equation*}
 	\end{dfn} 
	In the context of the Bayesian guessing game, $I(X,Y)$ is the ``value'' of being told the suit color, compared to having to play the game without that information. 
	From our calculations above, in the suit-guessing game, $I(X,Y) = H(X) - H(X|Y) = 2\log 2 - \log 2 = \log 2$. 

	Let's now express mutual information in two other ways. 
	Remarkably, these follow directly via simple algebra, but each identity provides a new way to think about the meaning of the mutual information. 
	\begin{thm}
		The mutual information may also be written as:
		\begin{align}
			I(X,Y) &= d(p_{X,Y}, p_X\otimes p_Y) \label{eq:mutual_independent}\\
				   &= \E_Y[d(p_{X|Y}, p_X)] \label{eq:expected_divergence}
		\end{align}
	\end{thm}
	
	We'll start by unpacking equation \eqref{eq:mutual_independent}, which expresses the mutual information $I(X,Y)$ as the divergence between the true joint distribution $p_{X,Y}$ and the product distribution $p_X\otimes p_Y$. 
	Recall that $p_X\otimes p_Y$ is what the distribution of $X$ and $Y$ would be, were they independent random variables. 
	Combining this observation with Gibbs' inequality, we obtain the following important facts: 
	\begin{cor}
		$I(X,Y) \geq 0$, with equality if and only if $X$ and $Y$ are independent.  
	\end{cor}
	So, $I(X,Y)$ is something like a super-charged correlation coefficient, in that it measures the degree of statistical dependence between $X$ and $Y$. However, the mutual information is more powerful than the correlation coefficient in two ways. 
	First, $I(X,Y)$ detects all kinds of statistical relationships, not just linear ones. 
	Second, while the correlation coefficient can vanish for dependent variables, this never happens for the mutual information. 
	Zero mutual information implies dependence, period. 
	As a quick illustration, it's not hard to calculate or intuit that if $X$ is the suit color and $Z$ is the numerical value of the card pulled, then $I(X,Z) = 0$. 
	Intuitively, if we were playing the suit-guessing game and I offered to tell you the card's face-value, you would be rightly annoyed. 
	That's an unhelpful (``uninformative'') offer, because the face-values and suit colors are independent.\footnote{So, why don't we just dispose of correlation coefficients and use $I(X,Y)$ instead? Well, correlation coefficients can be estimated from data relatively simply and are fairly robust to error. In contrast, $I(X,Y)$ requires that we have reasonably good estimates of the joint distribution $p_{X,Y}$, which is not usually available. Furthermore, it can be hard to distinguish $I(X,Y) = 10^{-6}$ from $I(X,Y) = 0$, and statistical tests of significance that address this problem are much more complex than those for correlation coefficients.} 

	Equation \eqref{eq:mutual_independent} has another useful consequence. 
	Since that formulation is symmetric in $X$ and $Y$, 
	\begin{cor}
		The mutual information is symmetric:
		\begin{equation*}
			I(X,Y) = I(Y,X)\;. 
		\end{equation*}
	\end{cor}
	
	Now let's unpack equation \eqref{eq:expected_divergence}. 
	One way to read this is as quantifying the danger of ignoring available information: $d(p_{X|Y = y}, p_X)$ is how surprised you would be if you ignored the information $Y=y$ and instead kept using $p_X$ as your belief. 
	If I told you that the deck contained only red cards, but you chose to ignore this and continue guessing $u = \left(\frac{1}{4},\frac{1}{4},\frac{1}{4},\frac{1}{4}\right)$ as your guess, you would be surprised to keep seeing red cards turn up draw after draw. 
	Formulation \eqref{eq:expected_divergence} expresses the mutual information as the expected surprise you would experience by ignoring your available side information $Y$, with the expectation taken over all the possible values the side information could assume. 
	While this formulation may seem much more opaque than \eqref{eq:mutual_independent}, it turns out to be remarkably useful when thinking geometrically, as it expresses the mutual information as the average ``distance'' between the marginal $p_X$ and the conditionals $p_{X|Y}$. 
	Pursuing this thought usefully expresses the mutual information as something like the ``moment of inertia'' for the joint distribution $p_{X,Y}$. 

\section{Why Information Shrinks}
	
	The famous 2nd Law of Thermodynamics states that, in a closed system, entropy increases. 
	The physicists' concept of entropy is closely related to but slightly different from the information theorist's concept, and we therefore won't make a direct attack on the 2nd Law in these notes. 
	However, there is a close analog of the 2nd Law that gives much of the flavor and can be formulated in information theoretic terms. 
	Whereas the 2nd Law states that entropy grows, the Data Processing Inequality states that information shrinks. 
	\begin{thm}[Data Processing Inequality]
		Let $X$ and $Y$ be random variables, and let $Z = g(Y)$, where $g$ is any function $g:\mathcal{Y} \rightarrow \mathcal{Z}$. Then, 
		\begin{equation*}
			I(X,Z) \leq I(X,Y). 		
		\end{equation*} 	
	\end{thm}
	This is not the most general possible form of the Data Processing Inequality, but it has the right flavor. 
	The meaning of this theorem is both ``obvious'' and striking in its generality.
	If you are using $Y$ to predict $X$, then any processing you do to $Y$ can only reduce your predictive power. 
	Data processing can enable tractable computations; reduce the impact of noise in your observations; and improve your visualizations. 
	The one thing it can't do is create information out of thin air. 
	No amount of processing is a substitute for having sufficient, salient data. 

	We'll pursue the proof the Data Processing Inequality, as the steps are quite enlightening. 
	First, we need the conditional mutual information: 
	\begin{dfn}[Conditional Mutual Information]
		The \textbf{conditional mutual information} of $X$ and $Y$ given $Z$ is
		\begin{equation*}
			I(X,Y|Z) = \sum_{z \in \mathcal{Z}} p_{Z}(z)d(p_{X,Y|Z = z},p_{X|Z = z}\otimes p_{Y|Z = z}).
		\end{equation*}
	\end{dfn}
	The divergence in the summand is naturally written $I(X,Y|Z = z)$, in which case we have $I(X,Y|Z) = \sum_{z \in \mathcal{Z}} p_Z(z) I(X,Y|Z = z)$, which has the form of an expectation of mutual informations conditioned on specific values of $Z$. 
	The conditional mutual information is naturally understood as the value of knowing $Y$ for the prediction of $X$, given that you already know $Z$. 
	Somewhat surprisingly, both of the cases $I(X,Y|Z) > I(X,Y)$ and $I(X,Y|Z) < I(X,Y)$ may hold; that is, knowing $Z$ can either increase or decrease the value of knowing $Y$ in the context of predicting $X$. 

	\begin{thm}[Chain Rule of Mutual Information]
		We have 
		\begin{equation*}
			I(X, (Y,Z)) = I(X, Z) + I(X, Y|Z).
		\end{equation*}
	\end{thm}
	\begin{remark}
		The notation $I(X, (Y,Z))$ refers to the (regular) mutual information between $X$ and the random variable $(Y,Z)$, which we can regard as a single random variable on the alphabet $\mathcal{Y} \times \mathcal{Z}$. 
	\end{remark}
	\begin{proof}
		We can compute directly, dividing up sums and remembering relations like $p_{X,Y,Z}(x,y,z) = p_{X,Y|Z}(x, y|z)p_{Z}(z)$. 
		Omitting some of the more tedious algebra,  
		\begin{align*}
			I(X,(Y,Z)) &= d(p_{X, Y, Z}, p_X\otimes p_{Y,Z}) \\ 
			           &= \sum_{x, z \in \mathcal{X} \times \mathcal{Z}} p_{X,Y|Z}(x,  z)  \log \frac{p_{X,Z}(x,z)}{p_X(x)p_Z(z) } + \\ &\quad  \sum_{x, y, z \in \mathcal{X} \times \mathcal{Y} \times \mathcal{Z}} p_Z(z)p_{X,Y|Z}(x, y| z) \log \frac{p_{X,Y|Z}(x,y|z)}{p_{Y|Z}(y|z)p_{X|Z}(x|z)} \\ 
			           &= I(X,Z) + I(X,Y|Z)\;,
		\end{align*}
		as was to be shown. 
	\end{proof}
	As always, the Chain Rule has a nice interpretation if you think about estimating $X$ by first learning $Z$, and then $Y$.   
	At the end of this process, you know both $Y$ and $Z$, and therefore have information $I(X,(Y,Z))$. 
	This information splits into two pieces: the information you gained when you learned $Z$, and the information you gained when you learned $Y$ after already knowing $Z$. 

	We are now ready to prove the Data Processing Inequality.\footnote{Proof borrowed from \url{http://www.cs.cmu.edu/~aarti/Class/10704/lec2-dataprocess.pdf}} 
	\begin{proof}
		Since $Z = g(Y)$, that is, is a function of $Y$ alone, we have that $Z \perp X | Y$, that is, given $Y$, $Z$ and $X$ are independent.\footnote{In fact, $Z \perp X|Y$ is often taken as the hypothesis of the Data Processing inequality rather than $Z = g(Y)$, as it is somewhat weaker and sufficient to prove the result.} 
		So, $I(X,Z|Y) = 0$. 
		On the other hand, using the chain rule in two ways, 
		\begin{align*}
			I(X, (Y,Z)) &= I(X,Z) + I(X,Y|Z) \\ 
			            &= I(X,Y) + I(X,Z|Y). 
		\end{align*}
		Since $I(X,Z|Y) = 0$ by our argument above, we obtain $I(X,Y) = I(X,Z) + I(X,Y|Z)$. Since $I(X,Y|Z)$ is nonnegative by Gibbs' inequality, we conclude that $I(X,Z) \leq I(X,Y)$, as was to be shown. 
	\end{proof}

	The Data Processing Inequality states that, in the absence of additional information sources, processing generically leaves you with less \emph{information} than you started. 
	The 2nd Law of Thermodynamics states that, in the absence of additional energy sources, the system dynamics leave you with less \emph{order} than you started. 
	These formulations suggest a natural parallel between the concepts of information and order, and therefore a natural parallel between the two theorems. 
	We'll close out this note with an extremely simplistic-yet-suggestive way to think about this. 

	Let $X_0$ and $Y_0$ each be random variables reflecting the possible locations and momenta of two particles at time $t = 0$. 
	We'll assume (a) that the particles don't interact, but that (b) the experimenter has placed the two particles very close to each other with similar momenta. 
	Thus, the initial configuration of the system is highly ordered, reflected by  $I(X_0, Y_0) > 0$. 
	If we knew $Y_0$, then we'd also significantly reduce our uncertainty about $X_0$. 
	How does this system evolve over time?
	We're assuming no interactions, so each of the particles evolve separately according to some short-timescale dynamics, which we can write $X_1 = g_x(X_0)$ and $Y_1 = g_y(Y_0)$. 
	Using the data processing inequality twice, we have 
	\begin{equation*}
		I(X_1, Y_1) \leq I(X_0, Y_1) \leq I(X_0, Y_0).
	\end{equation*}
	Thus, the dynamics tend to reduce information. 
	Of course, we can complicate this picture in various ways, by considering particle interactions or external potentials, either of which require a more sophisticated analysis. 
	The full 2nd Law, which is beyond the scope of these notes, is most appropriate for considering these cases. 

	\section{Some Further Reading}

		Those interested in these topics have many opportunities to explore them in more detail. 
		The below is a short list of some of the resources I have found most intriguing and useful, in addition to those cited in the introduction.

		\subsection*{Information Theory ``in General''}

		\begin{enumerate}
			\item Claude Shannon's original work \citep{Shannon1948}.
			\item Shannon's entertaining information-theoretic study of written English \citep{Shannon1951}.
			\item The text of \cite{Cover1991} is the standard modern overview of the field for both theorists and practitioners. 
			\item Colah's blog post ``Visual Information Theory'' at \url{http://colah.github.io/posts/2015-09-Visual-Information/} is both entertaining and extremely helpful for getting basic intuition around the relationship between entropy and communication. 
		\end{enumerate}

		\subsection*{Information Theory, Statistics, and Machine Learning}

			\begin{enumerate}
				\item An excellent and entertaining introduction to these topics is the already-mentioned \citet{MacKay2005}. 
				\item Those who want to further explore will likely enjoy \citet{Csiszzr2004}, but I would suggest doing this one after MacKay. 
				\item Readers interested in pursuing the Bayesian development of entropy much more deeply may enjoy \cite{Bernardo2008}, which provides an extremely thorough development of decision theory with a strong information-theoretic perspective.  
				\item The notes for the course ``Information Processing and Learning'' at Carnegie-Mellon's famous Machine Learning department are excellent and accessible; find them at \url{http://www.cs.cmu.edu/~aarti/Class/10704/lecs.html}
			\end{enumerate}

		\subsection*{Information Theory, Physics, and Biology}

		\begin{enumerate}
			\item Marc Harper has some intriguing papers in which he views biological evolutionary dynamics as learning processes through the framework of information theory; a few are \citep{Harper2009a,Harper2009}. 
			\item John Baez and his student Blake Pollard wrote a very nice and easy-reading review article of the role of information concepts in biological and chemical systems \citep{Baez2016}. 
			\item More generally, John Baez's blog is a treasure-trove of interesting vignettes and insights on the role that information plays in the physical and biological worlds: \url{https://johncarlosbaez.wordpress.com/category/information-and-entropy/}. For a more thoroughly worked-out connection between information dissipation and the Second Law of Thermodynamics, see this one: \url{https://johncarlosbaez.wordpress.com/2012/10/08/the-mathematical-origin-of-irreversibility/}. 
		\end{enumerate}

\bibliographystyle{apalike}

\end{document}